\documentclass{aa}
\usepackage{graphicx}
\usepackage{natbib}
\bibpunct{(}{)}{;}{a}{}{,}
\begin{document}
   \title{Optical monitoring of the gravitationally lensed quasar
Q2237+0305 from APO between June 1995 and January 1998\thanks{Based
on observations obtained with the Apache Point Observatory 3.5-meter
telescope, which is owned and operated by the Astrophysical Research
Consortium.}}

   \author{R. W. Schmidt\inst{1,2}$^{,}$\thanks{\protect\email{rschmidt@ast.cam.ac.uk}}
	\and T. Kundi\'c\inst{3} 
	\and U.-L. Pen\inst{4} 
	\and E. L. Turner\inst{5}
	\and J.  Wambsganss\inst{2}
%          \and ALL OBSERVERS (names provided by elt)
	\and L. E. Bergeron\inst{6}
	\and W. N. Colley\inst{7}
	\and C. Corson\inst{8}
	\and N. C. Hastings\inst{9}
	\and T. Hoyes\inst{10}
	\and D. C. Long\inst{9}
	\and K. A. Loomis\inst{9}
	\and S. Malhotra\inst{11}$^{,}$\thanks{Hubble Fellow}
	\and J. E. Rhoads\inst{6}
	\and K. Z. Stanek\inst{12}
          }

   \offprints{R.~W.~Schmidt}

   \date{Draft \today}% Received xxx 2001; accepted yyy 2001}

			%\fnmsep\thanks{email: rschmidt@ast.cam.ac.uk}, 
			%\fnmsep\thanks{email: jkw@astro.physik.uni-potsdam.de},

   \authorrunning{Schmidt et al.}
   \titlerunning{Optical monitoring of Q2237+0305 from June 1995 to
   January 1998}

   \institute{Institute of Astronomy,
              Madingley Road,
              Cambridge CB3 0HA,
              UK
%              \email{rschmidt@ast.cam.ac.uk}
              \and
		Universit\"at Potsdam,
              Institut f\"ur Physik,
              Am Neuen Palais 10,
              14469 Potsdam,
              Germany
%             \email{jkw@astro.physik.uni-potsdam.de}
              \and
              Renaissance Technologies,
              600 Route 25A,
              East Setauket,
              NY 11733,
              USA
%		\email{tomislav@rentec.com}
	\and
	Canadian Institute of Theoretical Astrophysics, 60 St. George 
	St., Toronto, M5S 3H8, Canada
%	\email{pen@cita.utoronto.ca}
              \and
              Princeton University Observatory,
              Peyton Hall,
              Princeton,
              NJ 08544,
              USA
%	      \email{elt@astro.princeton.edu}
	\and
	Space Telescope Science Institute, 3700 San Martin Drive,
              Baltimore, MD 21218, USA
	\and
	Lincoln Laboratory, Massachusetts Institute of Technology,
		244 Wood Street, Lexington, MA 02420, USA
	\and
	National Optical Astronomy Observatory, P.O. Box 26732,
		Tucson, AZ 85726, USA
	\and
	Apache Point Observatory, 2001 Apache Point Rd, Sunspot NM
              88349, USA
	\and
	Mesa Community College, 1833 West Southern Ave., Mesa,
              Arizona 85202-4866, USA
	\and
	Johns Hopkins University, Charles and 34th Street, Bloomberg
             Center, Baltimore, MD 21210, USA
	\and
              Harvard-Smithsonian Center for Astrophysics,
              60 Garden Street,
              Cambridge, MA 02138,
              USA
	}

   \abstract{We present a data set of images of the gravitationally
	lensed quasar Q2237+0305, that was obtained at the Apache
	Point Observatory (APO) between June 1995 and January
	1998. Although the images were taken under variable, often
	poor seeing conditions and with coarse pixel sampling,
	photometry is possible for the two brighter quasar images A
	and B with the help of exact quasar image positions from HST
	observations. We obtain a light curve with 73 data points for
	each of the images A and B. There is evidence for a long
	(${\mathrel{\hbox{\rlap{\hbox{\lower4pt\hbox{$\sim$}}}\hbox{$>$}}}}$\,100
	day) brightness peak in image A in 1996 with an amplitude of
	about 0.4 to 0.5 mag (relative to 1995), which indicates that
	microlensing has been taking place in the lensing
	galaxy. Image B does not vary much over the course of the
	observation period. The long, smooth variation of the light
	curve is similar to the results from the OGLE monitoring of
	the system~\citep{Wozniak00}.
   }

   \maketitle
%________________________________________________________________

\section{Introduction}

The quadruple quasar Q2237+0305 was discovered during the Center for
Astrophysics (CfA) redshift survey \citep{Huchra85}. In
high-resolution images of the system, four quasar images at a redshift
of $z = 1.695$ are seen in a cross-like geometry around the core of a
barred spiral galaxy with a redshift of $z=0.0394$.  Due to its
geometry this system is known as the Einstein cross
\citep{Schneider88,Yee88}.  The quasar images have a separation of
about $0.9$~arcsec from the galaxy centre.

Very quickly after its discovery it was realized that this system is
an ideal case for microlensing studies. The challenge in this system
is to measure the brightness of the four images individually with high
accuracy. In observations with seeing of larger than about an
arcsecond, it is very difficult to disentangle the four quasar images,
the galaxy core, and other features of the galaxy (such as the
bar-like structure that is situated across the galaxy centre, see
\citealt{Yee88}, or \citealt{Schmidt98}).

The record of semi-regular observations of this system began with the
announcement of the first microlensing event by \citet{Irwin89}.
\citet{Corrigan91} published the ``initial light curve'', that was
later augmented by other individual and systematical observations
\citep{Pen93,Houde94}. In \citet{Wambsganss92} it was emphasized that
without frequent sampling, it would be difficult to extract useful
information from the microlensing observations. The sample of early
observations with good seeing is rather heterogeneous regarding the
filters chosen, so that the interpretation is made difficult since the
filter differences have to be calibrated out.

\citet{Oestensen96} presented five years of observations of Q2237+0305
from the Nordic Optical Telescope (NOT). In all four images,
microlensing variations had then been detected. A rather striking drop
of about one magnitude in 1992 (with very large error bars) within
$\approx 20$ days had been found by \citet{Pen93} on the basis of
observations made at the Apache Point Observatory. In addition to this
photometric evidence, \cite{Lewis98} found spectroscopic signatures
for microlensing of the broad line region of this quasar.

Recently, the OGLE team has presented a light curve
\citep{Wozniak00,Wozniak00b} covering about 1200 days between 1997 and
2001.  Updated versions of this photometric data set can be looked at
at {\tt
www.astro.princeton.edu/\linebreak[0]\~{}ogle/\linebreak[0]ogle2/\linebreak[0]huchra.html}.
This data set provided a major step forward, and allows qualitatively
new approaches in the analysis of the light curves. The OGLE light
curves are very densely sampled and show amazing brightness variations
in all four quasar images with high amplitudes of more than one
magnitude.  Especially image C shows a dramatic brightness peak of
about 1.2 mag in 1999 that was resolved by the OGLE data in beautiful
detail.

In Sect. 2 our data set is described. In Sect. 3 we explain and
describe the details of the data reduction. In Sect. 4 we present our
results. We conclude in Sect. 5 with a discussion.

\section{Q2237+0305 at Apache Point Observatory}

We describe here the reduction and analysis of data that were taken
with the 3.5\,m telescope at Apache Point Observatory (APO) from June
1995 to January 1998 as part of the Princeton-APO lens monitoring
program. We restrict ourselves to the $r$-band data. The corresponding
CCD has a pixel size of $0.60$\arcsec. In Table~\ref{apo_log}, the
observation log is given.

\begin{table*}
\caption[Observation log: Q2237+0305 at APO, part I]{Observation log
of the $r$-band observations of Q2237+0305 at Apache Point Observatory
from 1995 to 1998. The table contains the date in yymmdd format, the
Julian date (-2449000), the total exposure time of all exposures in a
given night, the seeing as determined from reference star 3 in the
stacked images, and the magnitudes with 1\,$\sigma$ errors in
parentheses. }
\label{apo_log}
\vspace{0.25cm}
\begin{tabular}{lcccrr}
&&&&\multicolumn{2}{c}{} \\
&&total exposure\\
date & Julian & time per night & seeing & image A & image B \\
& date & (seconds) & (arcsec) & (mag) & (mag)\\
\hline
\\
950602 & 871 & 45 &  2.0  & 	17.57 (0.15) & 17.67 (0.17)\\
950606 & 875 & 300 & 2.2  & 	17.42 (0.20) & 17.80 (0.13)\\
950616 & 885 & 225 & 1.7  & 	17.30 (0.10) & 17.50 (0.10)\\
950620 & 889 & 195 & 2.1  & 	17.37 (0.12) & 17.30 (0.11)\\
950624 & 893 & 300 & 1.7  & 	17.20 (0.14) & 17.32 (0.13)\\
950625 & 894 & 150 & 1.9  & 	17.33 (0.16) & 17.53 (0.14)\\
950628 & 897 & 150 & 1.4  & 	17.32 (0.13) & 17.38 (0.10)\\
950702 & 901 & 195 & 2.0  & 	17.29 (0.15) & 17.47 (0.14)\\
950704 & 903 & 420 & 1.7  & 	17.32 (0.10) & 17.74 (0.12)\\
950710 & 909 & 390 & 1.7  & 	17.20 (0.13) & 17.53 (0.14)\\
950712 & 911 & 175 & 1.8  & 	16.92 (0.12) & 17.63 (0.16)\\
950722 & 921 & 525 & 1.6  & 	17.25 (0.12) & 17.68 (0.13)\\
950805 & 935 & 390 & 2.4  & 	17.31 (0.21) & 17.72 (0.17)\\
950819 & 949 & 60 &  2.0  & 	17.28 (0.17) & 17.45 (0.14)\\
950821 & 951 & 255 & 2.1  & 	17.20 (0.17) & 17.51 (0.16)\\
950823 & 953 & 180 & 2.2  & 	17.24 (0.15) & 17.75 (0.12)\\
950825 & 955 & 330 & 2.2  & 	17.36 (0.16) & 17.55 (0.15)\\
950827 & 957 & 375 & 2.2  & 	17.34 (0.19) & 17.44 (0.16)\\
950911 & 972 & 270 & 1.6  & 	17.24 (0.12) & 17.55 (0.13)\\
950923 & 984 & 270 & 1.9  & 	17.07 (0.12) & 17.73 (0.16)\\
950925 & 986 & 150 & 1.8  & 	17.02 (0.10) & 17.77 (0.16)\\
950929 & 990 & 510 & 1.8  & 	17.16 (0.12) & 17.53 (0.14)\\
951001 & 992 & 570 & 1.7  & 	17.25 (0.13) & 17.71 (0.14)\\
951003 & 994 & 540 & 1.9  & 	17.17 (0.12) & 17.51 (0.13)\\
951015 & 1006 & 300 &1.7  & 	17.08 (0.10) & 17.68 (0.13)\\
951017 & 1008 & 120 &1.6  & 	17.13 (0.09) & 17.88 (0.14)\\
       &      &     &     &             &            \\
960717 & 1282 & 180 & 1.4 & 	17.19 (0.10) & 17.94 (0.12)\\
960719 & 1284 & 210 & 1.4 & 	17.20 (0.12) & 17.66 (0.12)\\
960721 & 1286 & 270 & 1.3 & 	17.24 (0.08) & 17.93 (0.11)\\
960723 & 1288 & 210 & 1.4 & 	17.00 (0.10) & 17.79 (0.12)\\
960728 & 1293 & 150 & 1.3 & 	17.08 (0.10) & 17.76 (0.10)\\
960804 & 1300 & 240 & 1.6 & 	17.33 (0.10) & 17.71 (0.12)\\
960806 & 1302 & 330 & 1.6 & 	16.97 (0.09) & 17.71 (0.13)\\
960810 & 1306 & 270 & 1.4 & 	17.19 (0.09) & 17.80 (0.11)\\
960812 & 1308 & 270 & 1.4 & 	16.99 (0.09) & 17.86 (0.13)\\
960814 & 1310 & 360 & 1.8 & 	16.79 (0.13) & 17.64 (0.16)\\
960818 & 1314 & 270 & 1.4 & 	17.00 (0.11) & 17.72 (0.13)\\
960903 & 1330 & 180 & 1.4 & 	17.13 (0.09) & 17.66 (0.10)\\
960907 & 1334 & 90 &  1.5 & 	16.97 (0.09) & 17.75 (0.13)\\
960909 & 1336 & 90 &  1.8 & 	16.55 (0.11) & 18.04 (0.20)\\
960915 & 1342 & 180 & 2.5 & 	16.77 (0.19) & 17.95 (0.22)\\
960917 & 1344 & 180 & 1.9 & 	16.70 (0.14) & 17.73 (0.18)\\
960921 & 1348 & 90 &  1.4 & 	16.86 (0.10) & 17.63 (0.13)\\
960927 & 1354 & 270 & 1.9 & 	16.79 (0.12) & 17.82 (0.16)\\
960929 & 1356 & 120 & 2.0 & 	16.82 (0.15) & 17.85 (0.20)\\
961003 & 1360 & 90 &  2.8 & 	16.80 (0.20) & 18.12 (0.22)\\
961007 & 1364 & 180 & 2.3 & 	16.76 (0.23) & 17.89 (0.21)\\
961011 & 1368 & 120 & 1.3 & 	16.99 (0.09) & 17.72 (0.12)\\
961013 & 1370 & 285 & 1.3 & 	17.14 (0.07) & 17.84 (0.10)\\
961018 & 1375 & 60 &  1.9 & 	16.86 (0.14) & 17.48 (0.15)\\
\end{tabular}
\end{table*}

\begin{table*}
\addtocounter{table}{-1}
\caption[Observation log: Q2237+0305 at APO, part II]{(continued)}
\label{apo_log_II}
\vspace{0.25cm}
\begin{tabular}{lcccrr}
&&&&\multicolumn{2}{c}{}\\
&&total exposure\\
date & Julian & time per night & seeing & image A &
image B\\
& date & (seconds) & (arcsec) & (mag) & (mag)\\
\hline
\\
961030 & 1387 & 120 & 2.0 & 	16.68 (0.12) & 17.81 (0.20)\\
961101 & 1389 & 240 & 1.5 & 	16.86 (0.08) & 17.72 (0.14)\\
961103 & 1391 & 120 & 2.3 & 	16.92 (0.13) & 18.04 (0.21)\\
961107 & 1395 & 150 & 1.3 & 	16.98 (0.07) & 17.88 (0.11)\\
961109 & 1397 & 90 &  1.3 & 	17.14 (0.08) & 17.97 (0.13)\\
961111 & 1399 & 90 &  1.3 & 	16.94 (0.07) & 18.02 (0.14)\\
961120 & 1408 & 510 & 1.4 & 	17.00 (0.07) & 18.03 (0.13)\\
961124 & 1412 & 210 & 1.7 & 	17.17 (0.10) & 17.69 (0.10)\\
961126 & 1414 & 630 & 1.7 & 	16.94 (0.10) & 17.93 (0.15)\\
       &      &     &     &             &            \\
970715 & 1645 & 140 & 1.4 & 	17.05 (0.12) & 17.81 (0.15)\\
971018 & 1740 & 180 & 1.8 & 	17.21 (0.14) & 17.86 (0.16)\\
971026 & 1748 & 390 & 1.3 & 	17.28 (0.06) & 18.23 (0.10)\\
971030 & 1752 & 240 & 1.4 & 	17.34 (0.13) & 17.74 (0.13)\\
971101 & 1754 & 210 & 1.6 & 	17.45 (0.14) & 17.55 (0.12)\\
971103 & 1756 & 270 & 1.7 & 	17.25 (0.15) & 17.85 (0.16)\\
971105 & 1758 & 630 & 2.2 & 	17.52 (0.16) & 17.89 (0.14)\\
%971107 & 1760 & 120 & 0.9 & 	18.73(0.47) & 21.79(1.22)\\
971121 & 1774 & 220 & 1.7 & 	17.16 (0.11) & 17.97 (0.16)\\
971125 & 1778 & 630 & 1.9 & 	17.27 (0.20) & 17.74 (0.18)\\
971129 & 1782 & 450 & 1.3 & 	17.34 (0.09) & 17.88 (0.12)\\
980101 & 1815 & 30 &  1.7 & 	17.21 (0.10) & 17.85 (0.14)\\
980111 & 1825 & 60 &  2.0 & 	17.04 (0.15) & 17.64 (0.14)\\
\end{tabular}
\end{table*}

\begin{figure}
\centering
\resizebox{\hsize}{!}{
\includegraphics{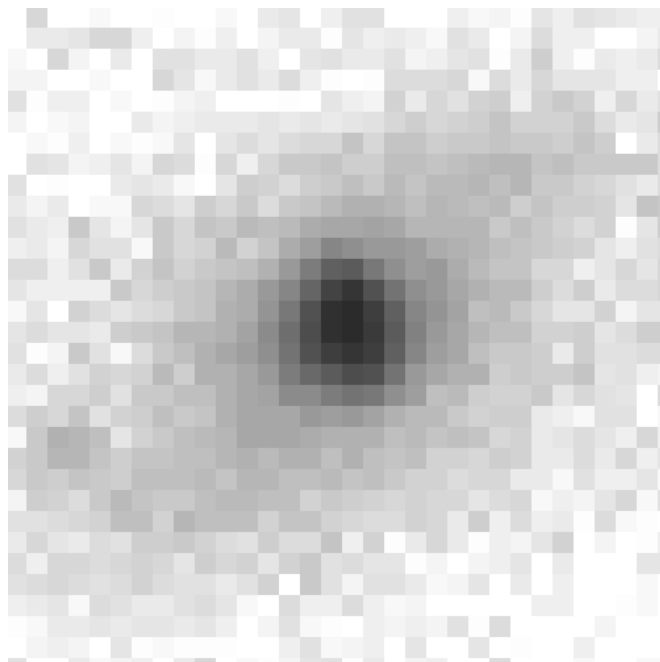}
}
\caption {Image of the central ($18.6\,$arcsec)$^2$ of the lensing
galaxy 2237+0305 in the $r$-band. The image was taken at APO on 20
November 1996 under 1.2\arcsec\, seeing. North is down and east is to
the left.}
\label{apoframe}
\end{figure}

The images were taken under variable, often poor seeing conditions
between 1.2{\arcsec} and several arcseconds. We have filtered out
frames with seeing worse than 2.4{\arcsec}. The distance between two
quasar images is only about three pixels. In these frames, the four
quasar images, the galaxy core, and the remaining light from the
galaxy are contained within only a dozen or so pixels, which is a
fairly coarse sampling compared to the number of sources in this
area. In Fig.~\ref{apoframe}, the centre of a frame obtained on 20
November 1996 under 1.2\arcsec\, seeing is shown. Except for the
overall central light concentration, the detailed quasar image- and
galaxy structure of this system is not visible.

However, we know the exact quasar image positions relative to the
galaxy centre from HST observation in the UV \citep{Blanton98} with
an accuracy of 5 mas. Moreover, a detailed model of the galaxy light
distribution is available \citep{Schmidt96} that was obtained from the
analysis of HST images as well. With the help of these strong
constraints, we are able to measure the brightness of the two brighter
quasar images A and B.

\section{Data reduction}
\label{2237reduction}

A total of 530 frames of Q2237+0305 were obtained at Apache Point
Observatory in the $r$-filter in 73 nights between June 1995 and
January 1998 (27 nights in 1995, 33 in 1996, 11 in 1997 and 2 in early
1998). The frames are processed using an automated procedure with the
following five steps of the data reduction:

\begin{enumerate}
\item Four reference stars in the vicinity of the galaxy were chosen
as reference points (see Fig.~\ref{apoframewithref*s}). We picked only
the brightest stars in order to be able to determine reliable
centroids. These reference stars are registered on each frame using an
automated routine.

\begin{figure}
\centering \resizebox{\hsize}{!}{
\includegraphics{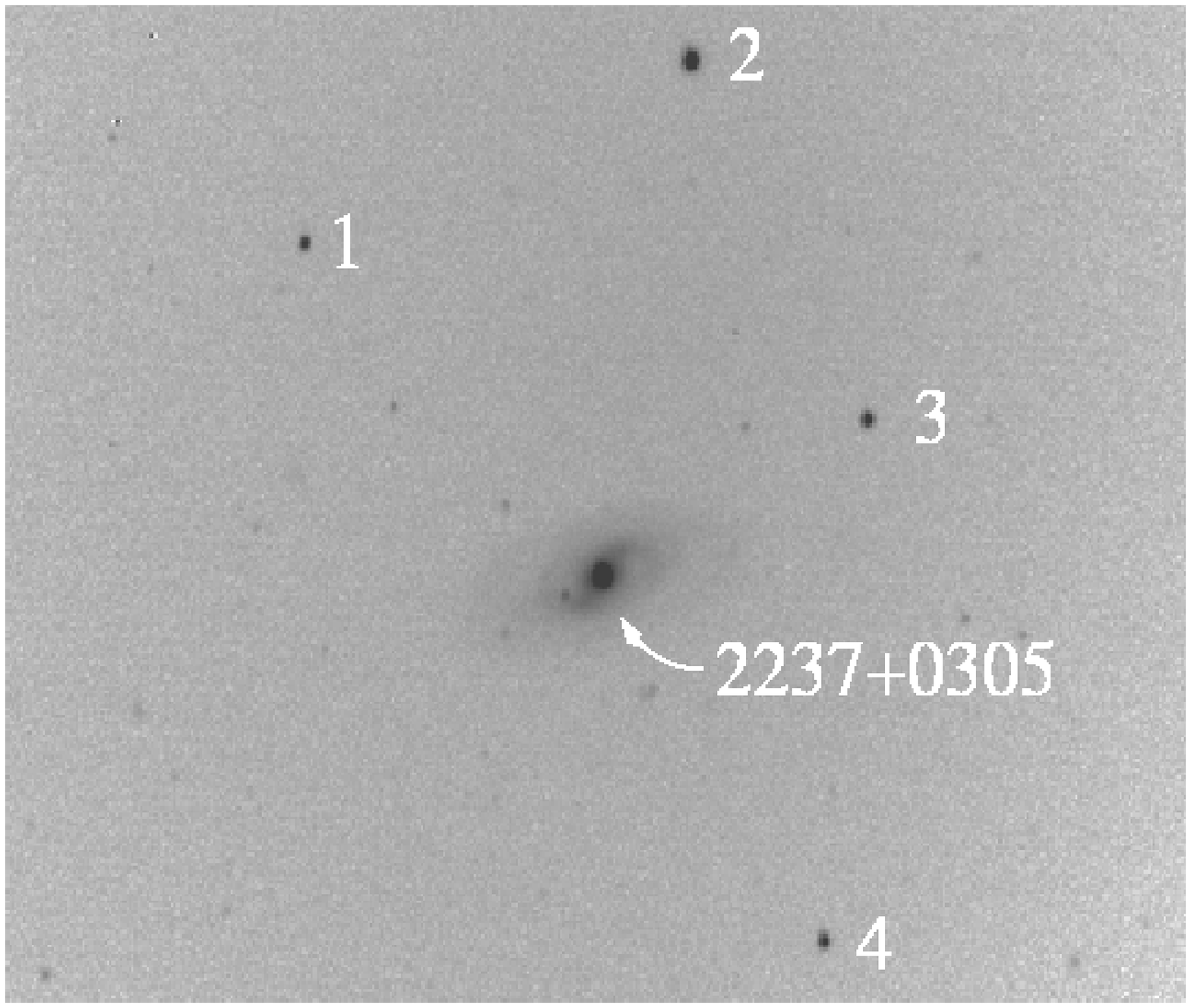}}
\caption{Image of the environs of the lensing galaxy 2237+0305 in the
$r$-band. This image was composed from 10 images taken at APO on 20
November 1996. The four reference stars are marked 1 to 4. The four
quasar images are situated in the core of the galaxy. The image size
is 4.2\,arcmin$\times$2.9\,arcmin. North is down and east is to the
left.}
\label{apoframewithref*s}
\end{figure}

\item Once the four reference stars are localized, the seeing is
determined by fitting a circular Gaussian brightness profile to the
reference stars with the IRAF task ``{\tt imexamine}''. These four
determinations are occasionally different from each other by up to ten
percent. We use the seeing of the star closest to the quasars
(star~\#3) as reference. The distances of stars \#2, \#3 and \#4 from
the centre of the galaxy are measured. These distances will be refined
in step~\ref{2ndfit}, but we need an approximate input to start the
quasar and galaxy fitting procedure in step~\ref{1stfit}.

The brightest reference star \#2 is used for the absolute flux
calibration. In Fig.~\ref{refstars}, the difference between the
measured magnitudes of stars \#2 and \#3 is depicted as a function of
time. The differences plotted in this figure scatter around a median
difference of $m_3-m_2=1.63$ mag, which is indicated by the dashed
line. The error bars are simply determined by Poisson statistics from
the flux counts. Except for a few outliers, the data are consistent
with our assumption that the two stars are not variable. If one or
both of the stars were variable, our photometry would still be
accurate to less than $0.02$ mag in $65\%$ of the cases, and
$0.04$\,mag in $95\%$ of the cases.  We use the $r-$magnitude
determined by \citet[][ their star $\alpha$]{Corrigan91} for star \#3
(m$_{3}=17.28$ mag) and the magnitude difference $m_3-m_2$ determined
above (with the described uncertainty) to obtain the $r-$band
magnitude of star \#2 as m$_{2}=15.65$ mag.

\begin{figure*}
\begin{center}
\includegraphics[width=17cm]{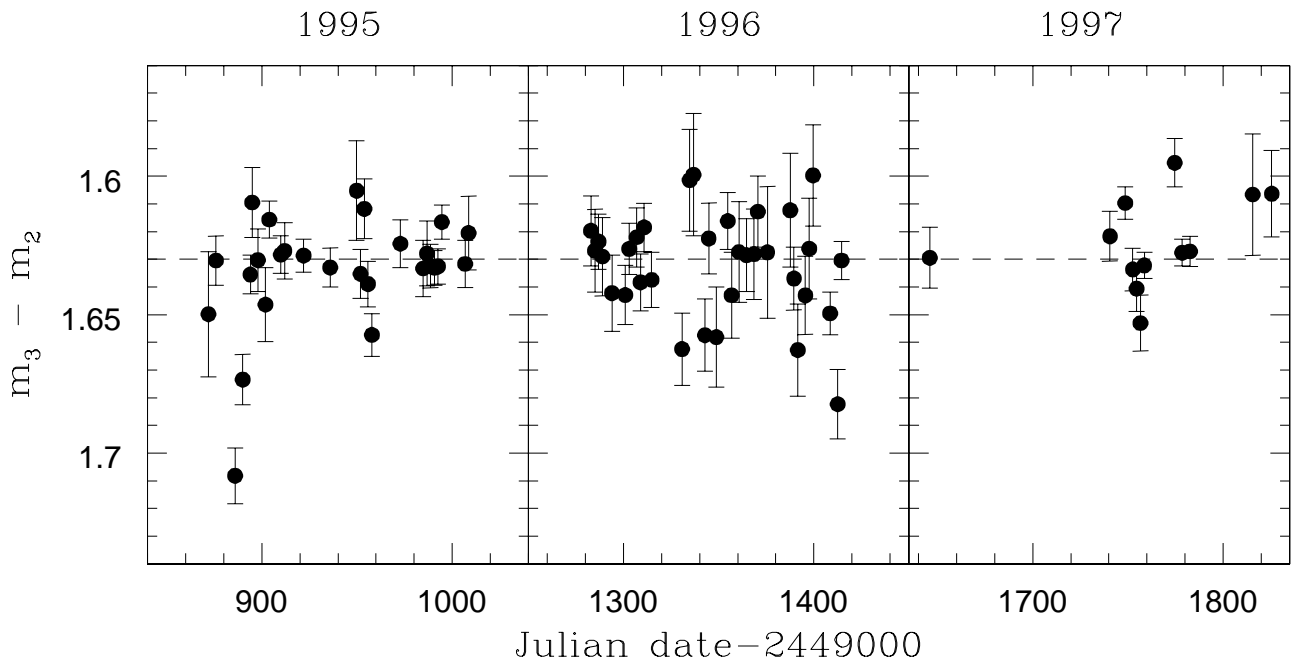}
\end{center}
\caption[Magnitude difference $m_3 -m_2$ between reference stars \#3
and \#2] {Magnitude difference $m_3 -m_2$ between reference stars \#3
and \#2 in the 73 combined frames.  The Poisson errors of the star
fluxes were added in quadrature. The median magnitude difference
1.63\,mag is shown with a dashed line.}
\label{refstars}
\end{figure*}

The frames of each night are added into a single image. The offsets
between different frames are calculated from all reference stars. The
frames are then aligned using the IRAF task ``{\tt imalign}''. In this
process, all frames are bilinearly interpolated to the positions of
one reference frame.

\item The light distribution of the galaxy is modelled numerically
using a model \citep{Schmidt96} that was obtained from the analysis of
an archival HST I-band image of the galaxy \citep{Westphal92}. The
model consists of a \citet{Devaucouleurs48} bulge and an exponential
disk. The De Vaucouleurs brightness profile $b(r)$ is given by (in
magnitudes per pixel):
\begin{equation}
b(r)=2.5\,{\rm log}\, i_0 - 3.33 \left[
\left(\frac{r}{r_{\rm bulge}} \right)^{1/4} -1 \right],
\end{equation}
where $i_0$ is the central intensity (in counts per pixel) and $r_{\rm
bulge}=(4.1\pm0.4)$~arcsec is the half-light radius. Since the galaxy
is inclined with respect to the line of sight, we use an elliptical
surface brightness distribution for the bulge with a constant axis
ratio of $0.69$ and a position angle, indicating the galaxy
inclination axis, of $77${\degr} (measured from north through
east). The exponential disk brightness profile in magnitudes per pixel
is given by
\begin{equation}
b(r)=2.5\,{\rm log}\, i_0-\Delta b_0
-\frac{r}{r_{\rm disk}}
\end{equation}
with $r_{\rm disk}=(11.3\pm1.2)$~arcsec and $\Delta
b_0=(1.0\pm0.1)$\,mag\,arcsec$^{-2}$. Due to the galaxy inclination,
the surface brightness distribution of the disk is seen with an axis
ratio of $0.5$. Although the Apache Point data were taken in the
$r$-band, we only left the amplitude $b_0$ a free parameter and, for
simplicity, kept the offset $\Delta b_0$ as measured using the HST
image. The galaxy light distribution was subsampled by a factor ten in
each direction before mapping it to the pixel grid of the
detector. The decomposition of the galaxy brightness profile into
different components is not unique. It is, for example, also possible
to use two exponential profiles, which could in general even be a
preferable solution \citep{Andredakis94}. However, any well-fitting
description of the light distribution in the inner 3 arcseconds of the
lensing galaxy will be good enough for us because the differences are
washed out by the coarse sampling of the data, the seeing at Apache
Point of more than an arcsecond, and noise.

In Fig.~\ref{apoframeminusgal}, the central pixels of the frame from
Fig.~\ref{apoframe} are shown after the galaxy model has been
subtracted. The remaining quasar images A and B (marked in the figure)
are seen. Images C and D cannot be detected with any significance.
Some very faint remaining structure is seen around the quasar images
that is due to the bar-like spiral arm structure that extends into the
galaxy centre \citep{Yee88,Schmidt98} and that had not been included
in the galaxy surface brightness model.

\begin{figure}
\begin{center}
\resizebox{8cm}{!}{\includegraphics{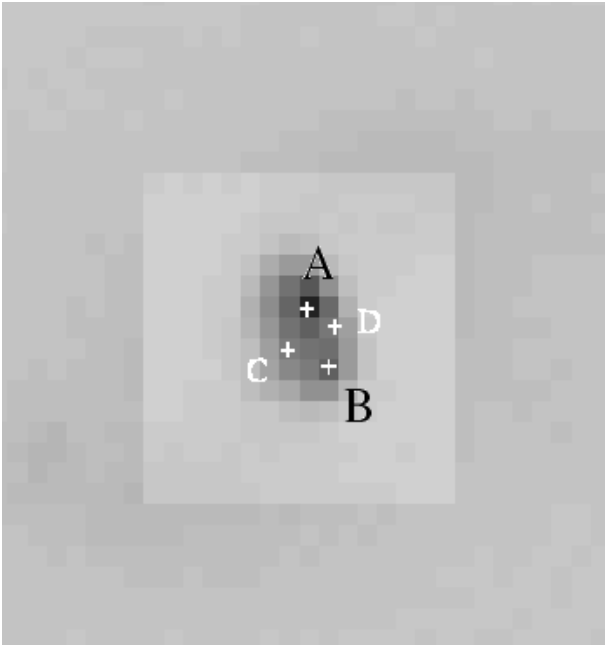}}
\end{center}
\caption[Image of the two quasar images A and B of Q2237+0305]{Image
of the central ($18.6\,$arcsec)$^2$ of the lensing galaxy
2237+0305 in the $r$-band.  The light from the galaxy has been
subtracted in the inner square with 9.6 arcsec (or 16 pixel) side
length.  This image was taken at APO on 20 November 1996 under
1.2{\arcsec}\, seeing. The raw image (before subtraction of the
galaxy) is shown in Fig.~\ref{apoframe}. The plus signs mark the
quasar image positions as determined from HST
\protect\citep{Blanton98}. The two point sources in the middle are the
quasar images A and B. Images C and D are not detected with any
significance because their amplitudes are not larger than the
noise. North is down and east is to the left.}
%
% Figure made with apo_pl_hst.fig --- circle positions calculated from
% the following command. Assuming 154.7 units/0.6 arcsec
%
% for circles:
%
% awk '{ex=257.8;cx=3950;cy=5290;for (i=2;i<=NF;i+=2) printf("1 3 0 1 7 7 50 0 -1 0.000 1 0.0000 %d %d 52 52 %d %d %d %d\n",int(cx+$(i-1)*ex),int(cy+$i*ex),int(cx+$(i-1)*ex),int(cy+$i*ex),int(cx+$(i-1)*ex)+52,int(cy+$i*ex))}' blanton_pos
%
% for pluses:
%
% awk '{semwid=52;ex=257.8;cx=3950;cy=5290;for (i=2;i<=NF;i+=2) printf("2 1 0 2 7 7 50 0 -1 0.000 0 0 -1 0 0 2\n%d %d %d %d\n2 1 0 2 7 7 50 0 -1 0.000 0 0 -1 0 0 2\n%d %d %d %d\n",int(cx+$(i-1)*ex-semwid),int(cy+$i*ex),int(cx+$(i-1)*ex+semwid),int(cy+$i*ex),int(cx+$(i-1)*ex),int(cy+$i*ex-semwid),int(cx+$(i-1)*ex),int(cy+$i*ex+semwid))}' blanton_pos
%
\label{apoframeminusgal}
\end{figure}

\item \label{1stfit} Given the seeing of our images, we consider the 5
mas accuracy of the \citet{Blanton98} HST coordinates of the four
quasar images relative to the galaxy centre as ``exact''. In the first
fitting run, the galaxy position and amplitude, as well as the four
quasar images are now fitted with the nonlinear minimization routine
AMOEBA \citep{Press92}. The quasar images are represented by four
circular Gaussian brightness profiles with a full width at half
maximum as determined from the seeing of the reference star. The
Gaussian profiles were subsampled by a factor ten in each direction
before mapping them to the pixel grid of the detector. The analytical
galaxy model is also convolved with this Gaussian seeing model. AMOEBA
is used to search for the best-fit solution in the parameter space of
galaxy position, galaxy amplitude, and quasar fluxes (as determined by
the amplitudes of the Gaussians) using a $\chi^2$ goodness of fit
estimator over the central $9\times 9$ pixels:
\begin{equation}
\label{chisqapo}
\chi^{2}=\frac{1}{80}\sum_{9\times 9\,{\rm inner}\,{\rm pixels}}
\frac{\left(i_{\rm model}-i_{\rm observed}\right)^{2}}{i_{\rm
observed}}.
\end{equation}

In order to constrain the parameter search to relevant parts of the
parameter space, we introduced additional constraints (priors) such
that trespassings across certain limit values were added quadratically
to eq.~(\ref{chisqapo}). This is used to enforce the constraints that
all fluxes are positive, and that the fluxes of image C and D are
smaller than the fluxes of A and~B. The latter assumption became wrong
in 1998 and 1999 where image C ``overtook'' image B during a
brightness peak beautifully resolved by OGLE.  However, our data
record stops in 1997, and until then there is no evidence for a
significant rise of either image C or D towards the fluxes of A or
B. As will be reported in Sect.~\ref{apo_results}, our flux estimates
have also been quantitatively confirmed at a few points by independent
observations.

\item \label{2ndfit} In step~\ref{1stfit} we have already produced a
light curve with flux estimates for all quasar images. In this final
step we use the results determined in step~\ref{1stfit} for all 73
nights and determine the median galaxy amplitude and the median
distances of the reference stars \#2, \#3, and \#4 from the centre of
the galaxy.

The fitting procedure is then repeated with a fixed value for the
galaxy amplitude. The galaxy position is also fixed by the distances
to the three reference stars \#2, \#3, and \#4. Thus, only the four
quasar fluxes are still free parameters. The distances of stars \#2,
\#3, and \#4 from the galaxy centre are
\begin{itemize}
\item[] $r_2$=(95.09$\pm$0.19) arcsec,
\item[] $r_3$=(64.55$\pm$0.12) arcsec,
\item[] $r_4$=(81.59$\pm$0.12) arcsec
\end{itemize}
(1$\sigma$ error bars). The result of the fitting process is a
best-fit model with parameters and uncertainties that can be estimated
from the depth and steepness of the $\chi^{2}$-surface dip
\citep{Press92}.

We estimate additional uncertainties of the fluxes by varying the
galaxy position within the range allowed by the separations of galaxy
centre and reference stars. This is done by using 10 random positions
for the galaxy allowed by the $1\sigma$-error bars of $r_2$, $r_3$ and
$r_4$ and by carrying through the whole analysis procedure. The error
bars given in Table~\ref{apo_log} and Fig.~\ref{2237curve} are
determined with this method.

\end{enumerate}

\section{Results}
\label{apo_results}

\begin{figure*}
\includegraphics[width=17cm]{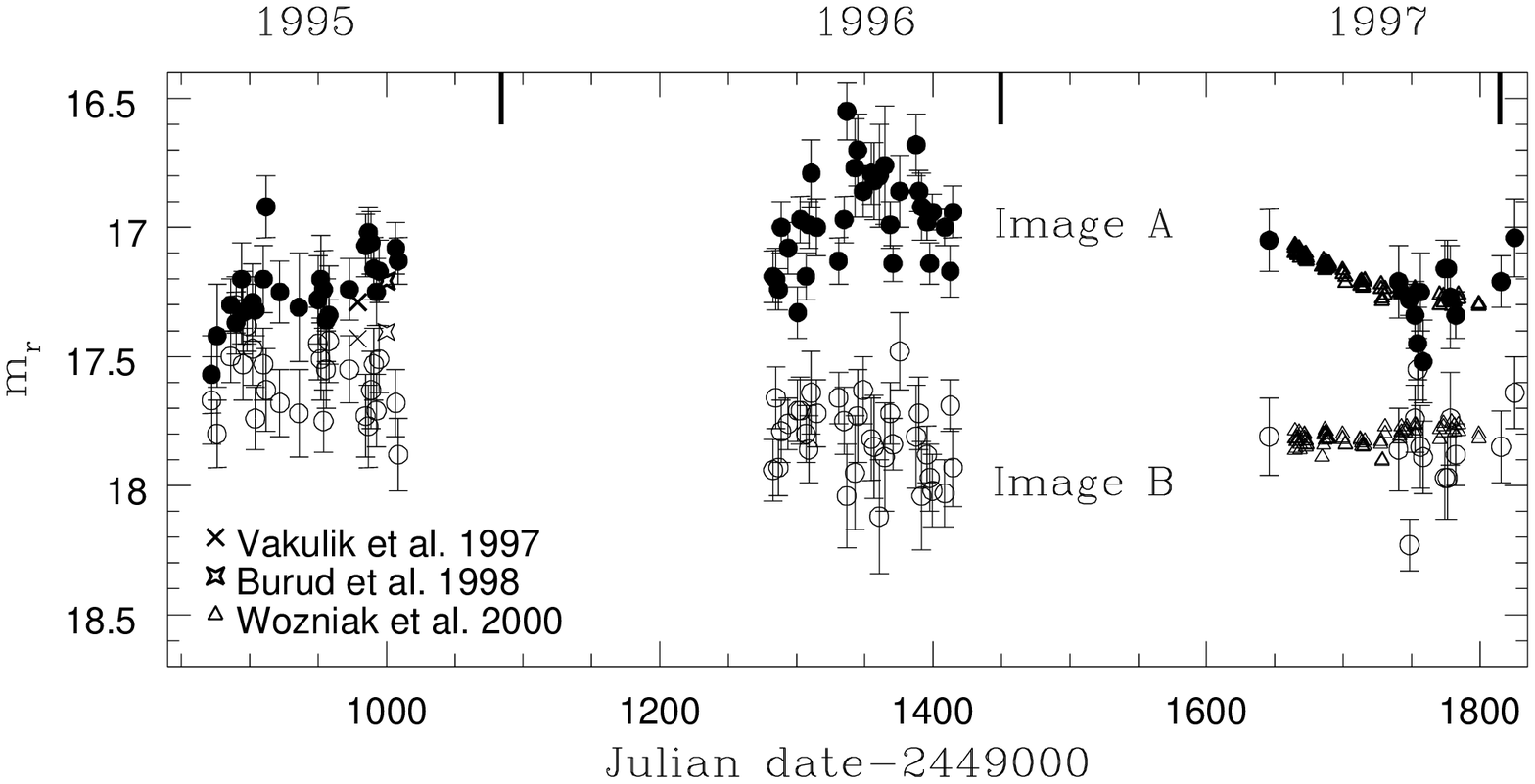}
\caption[Q2237+0305 light curve]{Final r-band light curve for images A
(filled circles) and B (open circles) of Q2237+0305 from June 1995 to
January 1998. In the text, the epochs are referred to as 1995 (left),
1996 (middle), and 1997 (right). The start of each year is denoted by
the thick markers at the top. The V-band data points by
\protect\citet[ crosses]{Vakulik97}, \protect\citet[ stars]{Burud98}
and \protect\citet[ triangles]{Wozniak00} are plotted with distinct
symbols in this figure because the associated error bars are only 0.05
mag or less. To plot these symbols we assume that the plotted relative
r-band magnitudes of image A can be obtained from the absolute V-band
magnitudes via $m_{\rm r}=m_{\rm V}-0.05$. For image B, we used
$m_{\rm r}=m_{\rm V}-0.01$ to correct for the colour difference (see
text).}
\label{2237curve}
\end{figure*}

We obtain r-band light curves for images A and B as shown in
Fig.~\ref{2237curve}, and detailed in Table~\ref{apo_log}. The fluxes
of images C and D are not measurable with our data since the images
are lost in the noise.

Despite the large error bars of the order of $\approx0.15$~mag for the
individual data points in Fig.~\ref{2237curve}, there is evidence for
a significant brightness peak in the light curve of image A in 1996
with an amplitude of about 0.4 to 0.5 mag (relative to
1995). Theoretical models of the system predict that intrinsic
fluctuations of the quasar appear in images A and B with a time-delay
of a few hours \citep{Schneider88,Wambsganss94}. Since image B does
not vary much -- there is some indication for a slight smooth decrease
-- the uncorrelated flux variation in image A can safely be
interpreted as microlensing in the lensing galaxy.

Since we had to carry out an involved data reduction process to obtain
photometry for the quasar images from the APO data, we used the V-band
data points by \citet[ crosses]{Vakulik97}, \citet[ stars]{Burud98}
and the OGLE group \citep[ triangles]{Wozniak00} for a
comparison. Since the APO data have been taken in the $r$-band, these
data points were plotted in Fig.~\ref{2237curve} by making the
assumption that the r-band magnitude of quasar image A on
JD-2449000=1800 is $m_{\rm r}=17.3$ mag. Assuming a simple additive
colour correction, we derive the relation $m_{\rm r}=m_{\rm V}-0.05$
from the OGLE data for image A. For image B, we used $m_{\rm r}=m_{\rm
V}-0.01$ in order to roughly account for the colour difference between
the two quasar images. This value was determined using the R-band
results from \citet{Vakulik97} (Maidanak Observatory, 17/19 September
1995, JD-2449000=978/980) and \citet{Burud98} (Nordic Optical
Telescope (NOT), 10/11 October 1995, JD-2449000=1000/1001).

Overall there is rather good agreement between our data and these
independent observations. In 1995, our results seem to favour a
slightly fainter image B. However, those of our data points that are
nearest in time to the independent measurements, agree within the error
bars. It is unclear whether this could be due to short-duration
small-scale microlensing, or a very small systematic offset. In 1996,
both the NOT ({\O}stensen 2001, priv. comm.)  and Maidanak Observatory
(Shalyapin 1999, priv. comm.)  observed that image A had become much
brighter than the other images.

We conclude that within the error bars we have quantitative
confirmation for our measurements in 1995 and 1997 for a few selected
epochs. In 1996 there is independent qualitative evidence for a
brightness rise of image A.

\section{Discussion}
\label{discussion}

We have presented monitoring data of Q2237+0305 over a period from
1995 to early 1998. Although the error bars on individual data points
are relatively large, the coverage of 73 nights clearly enables us to
see some significant trends in the microlensing behaviour. In
agreement with the results from the OGLE group
\citep{Wozniak00,Wozniak00b}, we find that Q2237+\linebreak[0]0305
shows large magnitude variations of several tenths of a magnitude on
timescales of less than hundred days.

It has become evident that the microlensing variations in
Q2237+\linebreak[0]0305 can happen rather smoothly over time spans of
several months to years. The \citet{Irwin89} event (also in its
interpretation by \citet{Racine92} as the first half of a double peak)
seems to be an example of a short-duration microlensing variation. The
\citet{Pen93} drop is an example of a high-magnitude short-term
process, but with very low signal-to-noise at that time.

All in all, it is fair to say that the observations of Q2237+0305 have
started to resolve microlensing variations in great detail, and they
look very much like the variations that were predicted already in the
early 80s for the microlensing effect. By filling up the gaps in the
light curves, long-term monitoring programs pave the way for the
statistical analysis of microlensing observations using, e.g.,
higher-order statistics of difference light curves \citep[e.g.,
][]{Kofman97}.

The OGLE light curve started where the data discussed in this paper
stopped and thus shows the continuation of our light curve. At Apache
Point, and at several other observatories (Maidanak Observatory, NOT),
data are still being taken. A new detector is being used at APO that
makes it possible to obtain much more accurate magnitude measurements
for the quasar images (see, for example, the images at {\tt
www.astro.princeton.edu/\linebreak[0]\~{}elt/2237.html}).

\begin{acknowledgements}

We thank the referee P. R. Wozniak for his detailed and careful report
and his very useful comments. We are grateful to Rainer K\"ohler for
providing his library of FITS C-routines. RWS and UP thank the
German-American Academic Council for sponsoring meetings in Aspen and
Ringberg during which part of this work was done. ELT acknowledges
support by the NSF grant AST98-02802. Part of this work was supported
by the German \emph{Deut\-sche For\-schungs\-ge\-mein\-schaft, DFG\/}
project number WA 1047/2-1.

\end{acknowledgements}

\end{document}